\RequirePackage{lineno}
\documentclass[aps,twocolumn,showpacs,byrevtex,prd,reprint]{revtex4-1}

\usepackage{graphicx}
\usepackage{dcolumn}
\usepackage{bm}
\usepackage{rotating}
\usepackage{epstopdf}
\usepackage{color}
\usepackage{verbatim} 
\usepackage{multirow}
\usepackage[abs]{overpic}
\usepackage{amsmath}

\newcommand{\PreserveBackslash}[1]{\let\temp=\\#1\let\\=\temp}
\newcolumntype{C}[1]{>{\PreserveBackslash\centering}p{#1}}
\newcolumntype{R}[1]{>{\PreserveBackslash\raggedleft}p{#1}}
\newcolumntype{L}[1]{>{\PreserveBackslash\raggedright}p{#1}}

\newcommand{\EE}{e^+e^-}

\newcommand{\too}{\rightarrow}

\begin{document}
\graphicspath{{figure/}}
\DeclareGraphicsExtensions{.eps,.png,.ps}

\title{\quad\\[0.0cm] \boldmath Search for vector charmonium(-like) states in $\EE \too \omega\chi_{cJ}$}

\author{Jielei Zhang}
\email{zhangjielei@ihep.ac.cn}
\author{Limin Yuan}
\affiliation{College of Physics and Electronic Engineering, Xinyang Normal University, Xinyang 464000, People's Republic of China}

\begin{abstract}
The cross sections of $\EE \too \omega\chi_{cJ(J=0,1,2)}$ have been measured by BESIII. We try to search for vector charmonium(-like) states $Y(4220)$, $Y(4360)$, $\psi(4415)$ and $Y(4660)$ in the $\EE \too \omega\chi_{cJ(J=0,1,2)}$ line shapes. The $\omega\chi_{c0}$ mainly comes from $Y(4220)$, $\omega\chi_{c1}$ mainly comes from $Y(4660)$ and $\omega\chi_{c2}$ mainly comes from $\psi(4415)$, maybe partly comes from $Y(4360)$ or $Y(4660)$. For the charmonium(-like) states that are not significant in the $\EE \too \omega\chi_{cJ(J=0,1,2)}$ line shape, we also give the $90\%$ confidence level upper limits on the electron partial width multiplied by branching fraction. These results are helpful to study the nature of charmonium(-like) states in this energy region.
\end{abstract}


\maketitle

In recent years, charmonium physics has gained renewed strong interest from both the theoretical and the experimental side, due to the observation of charmonium-like states, such as $X(3872)$~\cite{X3872aaa,CDF}, $Y(4260)$~\cite{Y4260,cleo,belleY4260}, $Y(4360)$~\cite{Y4360, Y4660} and $Y(4660)$~\cite{Y4660}. These states do not fit in the conventional charmonium spectroscopy, and could be exotic states that lie outside the quark model~\cite{exotic}. The $1^{--}$ $Y$-states are all observed in $\pi^{+}\pi^{-}J/\psi$ or $\pi^{+}\pi^{-}\psi(3686)$, while recently, one state (called $Y(4220)$) is observed in $\EE \too \omega \chi_{c0}$~\cite{omegachic,omegachic2}, and two states (called $Y(4220)$ and $Y(4390)$) are observed in $\EE \too \pi^{+}\pi^{-}h_{c}$~\cite{pipihc}. It indicates that the $Y$-states also can be searched by other charmonium transition decays. On the other hand, there are still some charmonium states predicted by the potential models which have not yet been observed experimentally, especially in the mass region higher than 4 GeV/$c^2$. The study of these $1^{--}$ $Y$-states is very helpful to clarify the missing predicted charmonium states in potential model. In all $Y$-states, maybe some are conventional charmonium. So it is important to confirm that which $Y$-states are charmonium and which $Y$-states are exotic states.

\begin{figure}[htbp]
\begin{center}
\includegraphics[width=0.43\textwidth]{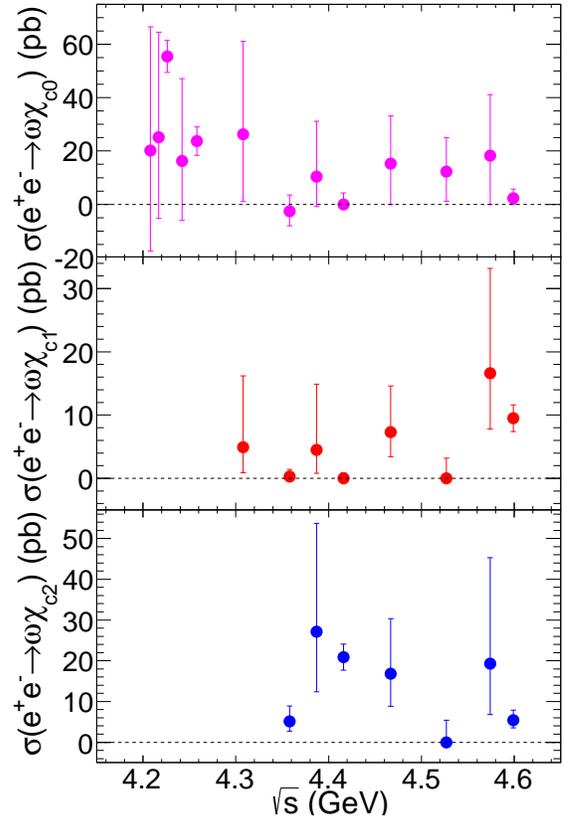}
\caption{Cross sections of $\EE \too \omega\chi_{cJ(J=0,1,2)}$ from BESIII. The top plot is for $\EE \too \omega\chi_{c0}$, the middle plot is for $\EE \too \omega\chi_{c1}$, and the bottom plot is for $\EE \too \omega\chi_{c2}$.}
\label{fig:crosssection}
\end{center}
\end{figure}

In all decay channels, the cross sections for $\EE \too \omega\chi_{cJ(J=0,1,2)}$ are relative large, so we can search for $Y$-states in $\omega\chi_{cJ(J=0,1,2)}$ line shape. The authors of Ref.~\cite{omegachic} perform a first search for the decay $\EE \too \omega\chi_{cJ(J=0,1,2)}$. The process $\EE \too \omega\chi_{c0}$ is observed around the center-of-mass energy $\sqrt{s}=4.23$ and $4.26$ GeV, while no significant $\omega\chi_{c1}$ and $\omega\chi_{c2}$ signals. Then, Ref.~\cite{omegachic2} also perform a search for $\EE \too \omega\chi_{cJ(J=0,1,2)}$ using the data from $\sqrt{s}=4.42$ to $4.6$ GeV. The decay $\EE \too \omega\chi_{c1}$ is observed around $\sqrt{s}=4.6$ GeV and decay $\EE \too \omega\chi_{c2}$ is observed around $\sqrt{s}=4.42$ GeV. The processes $\EE \too \omega\chi_{cJ(J=0,1,2)}$ are all observed, while the line shapes are different. Figure~\ref{fig:crosssection} shows the cross sections for $\EE \too \omega\chi_{cJ(J=0,1,2)}$ from BESIII for the center-of-mass energy from $\sqrt{s}=4.2$ to $4.6$ GeV. The different line shapes observed for $\omega\chi_{cJ(J=0,1,2)}$ might indicate that the production mechanisms are different, and that nearby resonances have different branching fractions to the $\omega\chi_{cJ(J=0,1,2)}$ decay modes.

Many theoretical papers have talked about the processes $\EE \too \omega\chi_{cJ(J=0,1,2)}$~\cite{the1,the2,the3,the4,the5}, so it is important to get the coupling strength between $\omega\chi_{cJ(J=0,1,2)}$ and different charmonium(-like) states, it can be helpful to develop the theory models. Above $4.2$ GeV, the all observed vector charmonium(-like) states are $Y(4220)$, $Y(4360)$, $\psi(4415)$ and $Y(4660)$.
In this paper, We try to search for these vector charmonium(-like) states in the $\EE \too \omega\chi_{cJ(J=0,1,2)}$ line shape. The $Y(4220)$ is above $\omega\chi_{c0}$ threshold, and the $Y(4360)$, $\psi(4415)$ and $Y(4660)$ are all above $\omega\chi_{c2}$ threshold.

From Fig.~\ref{fig:crosssection}, we can see there is an obvious structure around 4.23 GeV in the line shape of $\EE \too \omega\chi_{c0}$. Assuming that the $\omega\chi_{c0}$ signals come from a single resonance, we fit the cross section with a phase-space modified Breit-Wigner (BW) function; that is,
\begin{equation}
\begin{aligned}
\sigma(\sqrt{s})= & |BW(\sqrt{s})\sqrt{\frac{PS(\sqrt{s})}{PS(M)}}|^{2},
\end{aligned}
\end{equation}
where $PS(\sqrt{s})$ is the 2-body phase space factor, $BW(\sqrt{s})=\frac{\sqrt{12\pi\Gamma_{ee}\mathcal{B}(\omega\chi_{c0})\Gamma_{tot}}}{s-M^{2}+iM\Gamma_{tot}}$, is the BW function for a vector state, with mass $M$, total width $\Gamma_{tot}$, electron partial width $\Gamma_{ee}$, and the branching fraction to $\omega\chi_{c0}$, $\mathcal{B}(\omega\chi_{c0})$. From the fit, we can only extract $\Gamma_{ee}\mathcal{B}(\omega\chi_{c0})$.

\begin{figure}[htbp]
\begin{center}
\includegraphics[width=0.45\textwidth]{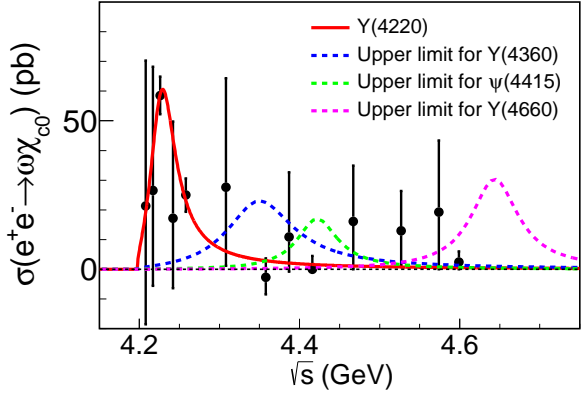}
\caption{Fit to the cross section of $\EE \too \omega\chi_{c0}$ from BESIII. The solid red curve shows the fit result using $Y(4220)$ structure, the dashed blue one is the $90\%$ C.L. upper limit for $Y(4360)$, the dashed green one is the $90\%$ C.L. upper limit for $\psi(4415)$, and the dashed purple one is the $90\%$ C.L. upper limit for $Y(4660)$.}
\label{fig:fit1}
\end{center}
\end{figure}

Figure~\ref{fig:fit1} shows the fit result. The fit results for the structure $Y(4220)$ are $M=(4226\pm8)$ MeV/$c^{2}$, $\Gamma=(39\pm12)$ MeV, and $\Gamma_{ee}\mathcal{B}(\omega\chi_{c0})=(2.8\pm0.5)$ eV. The goodness of the fit is $\chi^{2}/ndf=6.5/10$, corresponding to a confidence level of $77\%$. The mass and width are consistent with the state $Y(4220)$ found in $\EE \too \pi^{+}\pi^{-}h_{c}$~\cite{pipihc} and $\pi^{+}\pi^{-}J/\psi$~\cite{pipijpsi}. The cross sections for $\EE \too \omega\chi_{c0}$ around $\sqrt{s}=4.36, 4.42$ and $4.6$ GeV is close to 0, so the contributions from states $Y(4360)$, $\psi(4415)$ and $Y(4660)$ are small, we set $90\%$ confidence level (C.L.) upper limits for them.

Assuming that $\omega\chi_{c0}$ comes from two resonances $Y(4220)$ and $Y(4360)$, we fit the cross section with coherent sum of two constant width relativistic BW function; that is,
\begin{equation}
\begin{aligned}
\sigma(\sqrt{s})= & |BW_{1}(\sqrt{s})\sqrt{\frac{PS(\sqrt{s})}{PS(M_{1})}}+BW_{2}(\sqrt{s})\sqrt{\frac{PS(\sqrt{s})}{PS(M_{2})}}e^{i\phi_{1}}|^{2},
\end{aligned}
\end{equation}
where $\phi_{1}$ is relative phase, $BW_{1}$'s mass and width are fixed at the fit results for $Y(4220)$, and $BW_{2}$'s mass and width are fixed at the world average values~\cite{pdg} for $Y(4360)$. We use a least $\chi^{2}$ method to fit the cross section. The likelihood value can be got using the formula
\begin{equation}
\begin{aligned}
\mathcal{L}=e^{-\frac{1}{2}\chi^{2}}.
\end{aligned}
\end{equation}
We will calculate the $90\%$ C.L. upper limit on the electron partial width multiplied by branching fraction $\Gamma^{Y(4360)}_{ee}\mathcal{B}(Y(4360)\too\omega\chi_{c0})$ ($\Gamma\mathcal{B}$) for $Y(4360)$. The upper limit is determined by finding the value $(\Gamma\mathcal{B})^{\text{up}}$ such that $\int_{0}^{(\Gamma\mathcal{B})^{\text{up}}}\text{d}(\Gamma\mathcal{B})/\int_{0}^{\infty}\text{d}(\Gamma\mathcal{B})=0.90$, where $\mathcal{L}$ is the value of the likelihood as a function of $\Gamma\mathcal{B}$. From fit result, the $90\%$ C.L. upper limit for $Y(4360)$ is $\Gamma^{Y(4360)}_{ee}\mathcal{B}(Y(4360)\too\omega\chi_{c0})<3.0$ eV.

Using the same method, we also assume that $\omega\chi_{c0}$ comes from $Y(4220)$ and $\psi(4415)$, the upper limit for $\psi(4415)$ is determined to be $\Gamma^{\psi(4415)}_{ee}\mathcal{B}(\psi(4415)\too\omega\chi_{c0})<1.4$ eV. If we take $\Gamma(\psi(4415) \too \EE)=0.58$ keV~\cite{pdg}, we can obtain the $90\%$ C.L. upper limit for the branching fraction $\mathcal{B}(\psi(4415) \too \omega\chi_{c0})<2.4\times10^{-3}$. Assuming that $\omega\chi_{c0}$ comes from $Y(4220)$ and $Y(4660)$, the upper limit for $Y(4660)$ is $\Gamma^{Y(4660)}_{ee}\mathcal{B}(Y(4660)\too\omega\chi_{c0})<3.2$ eV. The upper limits for $Y(4360)$, $\psi(4415)$ and $Y(4660)$ are also shown in Fig.~\ref{fig:fit1}, and the results for $\EE \too \omega\chi_{c0}$ are listed in Table~\ref{tab:result1}.
\begin{table*}[htbp]
\begin{center}
\caption{ The fit results of the cross sections of $\EE \too \omega\chi_{cJ(J=0,1,2)}$, the upper limits are at $90\%$ C.L. }
\label{tab:result1}
\begin{tabular}{cccc}
  \hline
  \hline
  \qquad \qquad  & \qquad \qquad $\chi_{c0}$ \qquad \qquad \quad & \qquad \qquad $\chi_{c1}$ \qquad \qquad \quad  & \qquad \qquad  $\chi_{c2}$ \qquad \qquad \quad  \\
  \hline
  $\Gamma^{Y(4220)}_{ee}\mathcal{B}(Y(4220) \too \omega\chi_{cJ})$ (eV)  & $2.8\pm0.5$ & - & - \\
  $\Gamma^{Y(4360)}_{ee}\mathcal{B}(Y(4360) \too \omega\chi_{cJ})$ (eV) & $<3.0$ & $<0.5$ & $<3.0$ \\
  $\Gamma^{\psi(4415)}_{ee}\mathcal{B}(\psi(4415) \too \omega\chi_{cJ})$ (eV) & $<1.4$ & $<0.4$ & $2.1\pm0.3$ \\
  $\mathcal{B}(\psi(4415) \too \omega\chi_{cJ})$ ($\times10^{-3}$) & $<2.4$ & $<0.7$ & $3.6\pm0.5$ \\
  $\Gamma^{Y(4660)}_{ee}\mathcal{B}(Y(4660) \too \omega\chi_{cJ})$ (eV) & $<3.2$ & $2.9\pm0.6$ & $<4.7$ \\
  \hline
  \hline
\end{tabular}
\end{center}
\end{table*}

\begin{figure}[htbp]
\begin{center}
\includegraphics[width=0.45\textwidth]{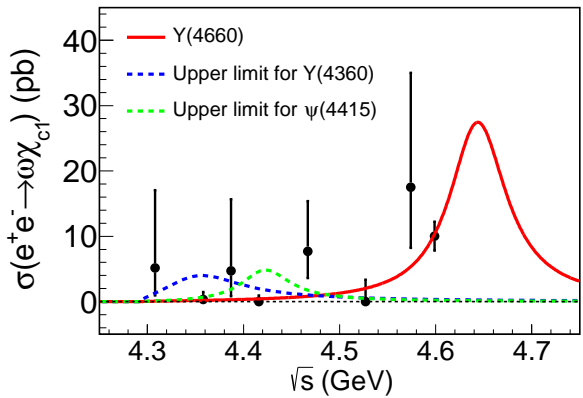}
\caption{Fit to the cross section of $\EE \too \omega\chi_{c1}$ from BESIII. The solid red curve shows the fit result using $Y(4660)$ structure, the dashed blue one is the $90\%$ C.L. upper limit for $Y(4360)$, and the dashed green one is the $90\%$ C.L. upper limit for $\psi(4415)$.}
\label{fig:fit2}
\end{center}
\end{figure}

From Fig.~\ref{fig:crosssection}, we can see there are obvious signals for $\EE \too \omega\chi_{c1}$ around $\sqrt{s}=4.6$ GeV, while no significant signals around $\sqrt{s}=4.36$ and $4.42$ GeV. The cross section of $\EE \too \omega\chi_{c1}$ seems to be rising near 4.6 GeV, it maybe from state $Y(4660)$. Assuming that the $\omega\chi_{c1}$ signals come from a single resonance $Y(4660)$, we fit the cross section with a phase-space modified BW function, the BW's mass and width are fixed at the world average values~\cite{pdg} for $Y(4660)$. Figure~\ref{fig:fit2} shows the fit result. The fit result for the structure $Y(4660)$ is $\Gamma^{Y(4660)}_{ee}\mathcal{B}(Y(4660)\too\omega\chi_{c1})=(2.9\pm0.6)$ eV. The goodness of the fit is $\chi^{2}/ndf=7.9/7$, corresponding to a confidence level of $34\%$.

Because the contributions from states $Y(4360)$ and $\psi(4415)$ are small, we also set $90\%$ C.L. upper limits for them. Assuming that $\omega\chi_{c1}$ comes from $Y(4660)$ and $Y(4360)$, the upper limit for $Y(4360)$ is $\Gamma^{Y(4360)}_{ee}\mathcal{B}(Y(4360)\too\omega\chi_{c1})<0.5$ eV. We also assume that $\omega\chi_{c1}$ comes from $Y(4660)$ and $\psi(4415)$, the upper limit for $\psi(4415)$ is determined to be $\Gamma^{\psi(4415)}_{ee}\mathcal{B}(\psi(4415)\too\omega\chi_{c1})<0.4$ eV. If we take $\Gamma(\psi(4415) \too \EE)=0.58$ keV~\cite{pdg}, we can obtain the $90\%$ C.L. upper limit for the branching fraction $\mathcal{B}(\psi(4415) \too \omega\chi_{c1})<0.7\times10^{-3}$. The upper limits for $Y(4360)$ and $\psi(4415)$ are also shown in Fig.~\ref{fig:fit2}, and the results for $\EE \too \omega\chi_{c1}$ are also listed in Table~\ref{tab:result1}.

\begin{figure}[htbp]
\begin{center}
\includegraphics[width=0.45\textwidth]{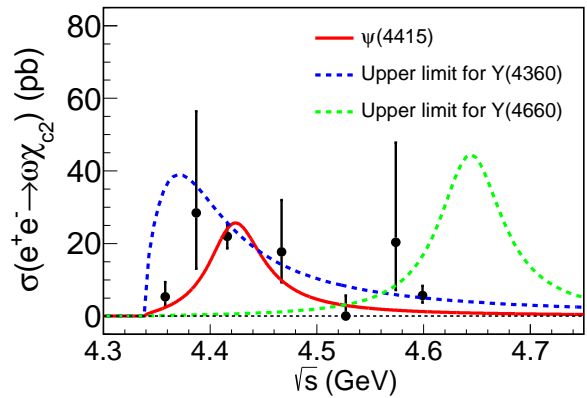}
\caption{Fit to the cross section of $\EE \too \omega\chi_{c2}$ from BESIII. The solid red curve shows the fit result using $\psi(4415)$ structure, the dashed blue one is the $90\%$ C.L. upper limit for $Y(4360)$, and the dashed green one is the $90\%$ C.L. upper limit for $Y(4660)$.}
\label{fig:fit3}
\end{center}
\end{figure}

From Fig.~\ref{fig:crosssection}, we can see there are obvious signals for $\EE \too \omega\chi_{c2}$ around $\sqrt{s}=4.42$ GeV, while signals are not significant around $\sqrt{s}=4.36$ and $4.6$ GeV. It seems there is an enhancement around 4.42 GeV, $\omega\chi_{c2}$ maybe from state $\psi(4415)$. Assuming that the $\omega\chi_{c2}$ signals come from a single resonance $\psi(4415)$, we fit the cross section with a phase-space modified BW function, the BW's mass and width are fixed at the world average values~\cite{pdg} for $\psi(4415)$. Figure~\ref{fig:fit3} shows the fit result. The fit result for the structure $\psi(4415)$ is $\Gamma^{\psi(4415)}_{ee}\mathcal{B}(\psi(4415)\too\omega\chi_{c2})=(2.1\pm0.3)$ eV. If we take $\Gamma(\psi(4415) \too \EE)=0.58$ keV~\cite{pdg}, we can obtain $\mathcal{B}(\psi(4415) \too \omega\chi_{c2})=(3.6\pm0.5)\times10^{-3}$. The goodness of the fit is $\chi^{2}/ndf=11.3/6$, corresponding to a confidence level of $8\%$.

Because the signals are not significant around $\sqrt{s}=4.36$ and $4.6$ GeV, we also set $90\%$ C.L. upper limits for $Y(4360)$ and $Y(4660)$. Assuming that $\omega\chi_{c2}$ comes from $\psi(4415)$ and $Y(4360)$, the upper limit for $Y(4360)$ is $\Gamma^{Y(4360)}_{ee}\mathcal{B}(Y(4360)\too\omega\chi_{c2})<3.0$ eV. We also assume that $\omega\chi_{c2}$ comes from $\psi(4415)$ and $Y(4660)$, the upper limit for $Y(4660)$ is determined to be $\Gamma^{Y(4660)}_{ee}\mathcal{B}(Y(4660)\too\omega\chi_{c2})<4.7$ eV. The upper limits for $Y(4360)$ and $Y(4660)$ are also shown in Fig.~\ref{fig:fit3}, and the results for $\EE \too \omega\chi_{c2}$ are also listed in Table~\ref{tab:result1}.

\begin{table*}[htbp]
\begin{center}
\caption{ The fit results of the cross section of $\EE \too \omega\chi_{c2}$. One is the results using $\psi(4415)$ and $Y(4360)$ to fit, and the other one is the results using $\psi(4415)$ and $Y(4660)$ to fit. }
\label{tab:result2}
\begin{tabular}{ccccc}
  \hline
  \hline
          &  \multicolumn{2}{c}{$\psi(4415)+Y(4360)$} & \multicolumn{2}{c}{$\psi(4415)+Y(4660)$}  \\
  \quad & \quad SolutionI \quad \quad & \quad SolutionII \quad \quad & \quad SolutionI \quad \quad & \quad SolutionII \quad \quad \\
  \hline
  $\Gamma^{\psi(4415)}_{ee}\mathcal{B}(\psi(4415) \too \omega\chi_{c2})$ (eV) & $1.6\pm1.0$ & $7.1\pm1.2$ & $1.8\pm0.3$ & $2.5\pm0.4$ \\
  $\mathcal{B}(\psi(4415) \too \omega\chi_{cJ})$ ($\times10^{-3}$) & $2.8\pm1.7$ & $12.2\pm2.1$ & $3.1\pm0.5$ & $4.3\pm0.7$ \\
  $\Gamma^{Y(4360)/Y(4660)}_{ee}\mathcal{B}(Y(4360)/Y(4660) \too \omega\chi_{c2})$ (eV) & $0.6\pm0.4$ & $2.2\pm0.8$ & $1.4\pm2.0$ & $3.0\pm2.2$ \\
  $\phi_{1}$ & $0.53\pm0.60$ & $2.16\pm0.15$ & $0.75\pm1.47$ & $-1.58\pm0.88$ \\
  \hline
  \hline
\end{tabular}
\end{center}
\end{table*}

\begin{figure}[htbp]
\begin{center}
\includegraphics[width=0.45\textwidth]{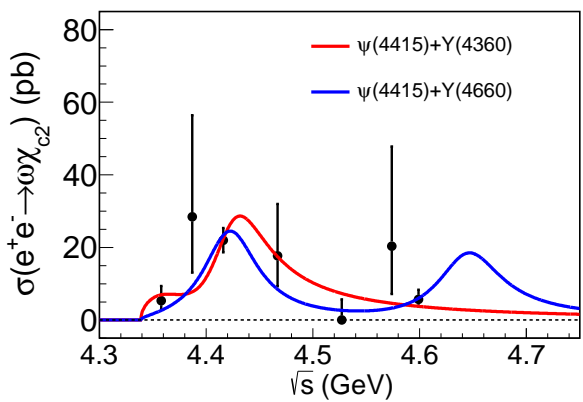}
\caption{Fit to the cross section of $\EE \too \omega\chi_{c2}$ from BESIII. The solid red curve shows the fit result using $\psi(4415)$ and $Y(4360)$, and solid blue curve shows the fit result using $\psi(4415)$ and $Y(4660)$.}
\label{fig:fit4}
\end{center}
\end{figure}

If we only use a $\psi(4415)$ to fit the cross section of $\EE \too \omega\chi_{c2}$, the goodness of the fit is $\chi^{2}/ndf=11.3/6$. The goodness of the fit is relatively large, it indicates maybe there are contributions from other charmonium(-like) states. Assuming that $\omega\chi_{c2}$ comes from two resonances $\psi(4415)$ and $Y(4360)$, we fit the cross section with coherent sum of two constant width relativistic BW function. Figure~\ref{fig:fit4} shows the fit result. There are two solutions with same fit quality, the results are listed in Table~\ref{tab:result2}. The goodness of the fit is $\chi^{2}/ndf=5.9/4$, corresponding to a confidence level of $21\%$. Comparing the $\chi^{2}$s change and taking into the change of the number of degree of freedom, the statistical significance of the $Y(4360)$ resonance is $1.8\sigma$. We also try to assume that $\omega\chi_{c2}$ comes from two resonances $\psi(4415)$ and $Y(4660)$, the fit result is also shown in Fig.~\ref{fig:fit4}. There are two solutions with same fit quality, the results are also listed in Table~\ref{tab:result2}. The goodness of the fit is $\chi^{2}/ndf=5.9/4$, corresponding to a confidence level of $21\%$. Comparing the $\chi^{2}$s change and taking into the change of the number of degree of freedom, the statistical significance of the $Y(4660)$ resonance is $1.8\sigma$. The goodness of the fits are same with the two assumptions. With more data sample in the future, especially the data above 4.6 GeV, it can be used to distinguish which hypothesis is reasonable.

In summary, we try to search for vector charmonium(-like) states $Y(4220)$, $Y(4360)$, $\psi(4415)$ and $Y(4660)$ in the $\EE \too \omega\chi_{cJ(J=0,1,2)}$ line shapes. The $\omega\chi_{c0}$ mainly comes from $Y(4220)$, $\omega\chi_{c1}$ maybe mainly comes from $Y(4660)$ and $\omega\chi_{c2}$ mainly comes from $\psi(4415)$. More data samples are need to confirm these assumptions, and it is very important to confirm the structure above 4.6 GeV in $\EE\too\omega\chi_{c1}$. For the charmonium(-like) states that are not significant in the $\EE \too \omega\chi_{cJ(J=0,1,2)}$ line shape, we also give the $90\%$ C.L. upper limits on the electron partial width multiplied by branching fraction. The results are listed in Table~\ref{tab:result1}. We also try to use $\psi(4415)$ and $Y(4360)/Y(4660)$ to fit the cross section of $\EE \too \omega\chi_{c2}$, the results are listed in Table~\ref{tab:result2}. It will be helpful to study the nature of charmonium(-like) states. More high precision measurements around this energy region are desired to better understand these results, this can be achieved in BESIII and BelleII experiments in the further.


\section*{Acknowledgement}
This work is supported by Nanhu Scholars Program for Young Scholars of Xinyang Normal University and Scientific Research Foundation of Graduate School of Xinyang Normal University.


\begin{thebibliography}{**}

\bibitem{X3872aaa} S.~K.~Choi {\it et al.} [Belle Collaboration], Phys.\ Rev.\ Lett.\  {\bf 91}, 262001 (2003).

\bibitem{CDF} D.~Acosta {\it et al.} [CDF Collaboration], Phys.\ Rev.\ Lett.\  {\bf 93}, 072001 (2004).

\bibitem{Y4260} B.~Aubert {\it et al.} [BaBar Collaboration], Phys.\ Rev.\ Lett.\  {\bf 95}, 142001 (2005).

\bibitem{cleo} T.~E.~Coan {\it et al.} [CLEO Collaboration], Phys.\ Rev.\ Lett.\  {\bf 96}, 162003 (2006).

\bibitem{belleY4260} C.~Z.~Yuan {\it et al.} [Belle Collaboration], Phys.\ Rev.\ Lett.\  {\bf 99}, 182004 (2007).

\bibitem{Y4360} B.~Aubert {\it et al.} [BaBar Collaboration], Phys.\ Rev.\ Lett.\  {\bf 98}, 212001 (2007).

\bibitem{Y4660} X.~L.~Wang {\it et al.} [Belle Collaboration], Phys.\ Rev.\ Lett.\  {\bf 99}, 142002 (2007).

\bibitem{exotic} N.~Brambilla, S.~Eidelman, B.~K.~Heltsley, R.~Vogt, G.~T.~Bodwin, E.~Eichten, A.~D.~Frawley and A.~B.~Meyer {\it et al.}, Eur.\ Phys.\ J.\ C {\bf 71}, 1534 (2011).

\bibitem{omegachic} M.~Ablikim {\it et al.} [BESIII Collaboration], Phys.\ Rev.\ Lett.\  {\bf 114}, 092003 (2015).

\bibitem{omegachic2} M.~Ablikim {\it et al.} [BESIII Collaboration], Phys.\ Rev.\ D {\bf 93}, 011102 (2016).

\bibitem{pipihc} M.~Ablikim {\it et al.} [BESIII Collaboration], Phys.\ Rev.\ Lett.\  {\bf 118}, 092002 (2017).

\bibitem{the1} X.~Li and M.~B.~Voloshin, Phys.\ Rev.\ D {\bf 91}, 034004 (2015).

\bibitem{the2} D.~Y.~Chen, X.~Liu and T.~Matsuki, Phys.\ Rev.\ D {\bf 91}, 094023 (2015).

\bibitem{the3} L.~Ma, X.~H.~Liu, X.~Liu and S.~L.~Zhu, Phys.\ Rev.\ D {\bf 91}, 034032 (2015).

\bibitem{the4} R.~Faccini, G.~Filaci, A.~L.~Guerrieri, A.~Pilloni and A.~D.~Polosa, Phys.\ Rev.\ D {\bf 91}, 117501 (2015).

\bibitem{the5} M.~Cleven and Q.~Zhao, Phys.\ Lett.\ B {\bf 768}, 52 (2017).

\bibitem{pipijpsi} M.~Ablikim {\it et al.} [BESIII Collaboration], Phys.\ Rev.\ Lett.\  {\bf 118}, 092001 (2017).

\bibitem{pdg} C.~Patrignani {\it et al.} [Particle Data Group], Chin.\ Phys.\ C {\bf 40}, 100001 (2016).

\end{thebibliography}
\end{document}